\documentclass[aps,prb,twocolumn,tightenlines,superscriptaddress,amsmath,amssymb,longbibliography]{revtex4-1}
\pdfoutput=1
\usepackage{graphicx}
\usepackage{color}
\usepackage{bm}
\usepackage{amsmath}
\usepackage{amsfonts}
\usepackage{amssymb}

\usepackage{comment}
\usepackage{graphicx,ams math,epsfig,epstopdf}
\usepackage{dcolumn}
\usepackage{bm}
\usepackage{hyperref}
\usepackage{tabularx}
\usepackage{multirow}  
\hypersetup{
	colorlinks = True,
	urlcolor=blue,
	linkcolor = blue,
	citecolor= blue
}
\usepackage{grffile}
\usepackage{color}
\usepackage[english]{babel}

 \newcommand{\be}{\begin{equation}}
 \newcommand{\ee}{\end{equation}}
\newcommand{\bea}{\begin{eqnarray}}
\newcommand{\eea}{\end{eqnarray}}
\newcommand{\ba}{\begin{eqnarray*}}
\newcommand{\ea}{\end{eqnarray*}}

\newcommand{\bx}{\mathbf{x}}
\newcommand{\by}{\mathbf{y}}

\usepackage{multirow}

\begin{document}




\title{Quantum Monte Carlo tunneling from quantum chemistry to quantum annealing} 

\author{Guglielmo Mazzola}
 \email[]{gmazzola@phys.ethz.ch}
 \affiliation{Theoretische Physik, ETH Zurich, 8093 Zurich, Switzerland}

 \author{Vadim N. Smelyanskiy}
 \affiliation{Google, Venice, CA 90291, USA}

\author{Matthias Troyer}
 \affiliation{Theoretische Physik, ETH Zurich, 8093 Zurich, Switzerland}
\affiliation{Quantum Architectures and Computation Group, Station Q, Microsoft Research,
Redmond, WA 98052, USA}
\date{\today}

\begin{abstract}
Quantum Tunneling is ubiquitous across different fields, from quantum chemical reactions, and magnetic materials to quantum simulators and quantum computers. While simulating the real-time quantum  dynamics of tunneling is infeasible for high-dimensional systems, quantum tunneling also shows up in quantum Monte Carlo (QMC) simulations that scale polynomially with system size.
Here we extend a recent results obtained for quantum spin models {[{Phys. Rev. Lett.} {\bf 117}, 180402 (2016)]}, and study high-dimensional continuos variable models for proton transfer reactions. We demonstrate that QMC simulations efficiently recover ground state tunneling rates due to the existence of an instanton path, which always connects the reactant state with the product. 
We discuss the implications of our results in the context of quantum chemical reactions and quantum annealing, where quantum tunneling is expected to be a valuable resource for solving combinatorial optimization problems.

\end{abstract}

\maketitle

\section{Introduction}

Quantum mechanical tunneling (QMT)  plays a fundamental role  in a broad range of disciplines, from chemistry and physics to quantum computing.
QMT can be observed in chemical reactions\cite{nagel2006tunneling,bell2013tunnel,zuev2003carbon,nakamura2013quantum} and affects the description of water and related aqueous system  at room temperature.\cite{ceriotti2016nuclear,richardson2016concerted}
It is essential for understanding -- even at the qualitative level -- the phase diagrams of correlated materials, such as dense hydrogen, which is  the simplest condensed matter system.\cite{bonev2004quantum,pickard_structure_2007,dalladay2016evidence,mazzola2017accelerating} 

QMT can also been engineered in quantum annealers,\cite{johnson2011quantum,dwave2x} to solve optimization problems using quantum effects.\cite{farhi2000quantum,PhysRevE.58.5355,Farhi20042001,RevModPhys.80.1061,crosson2016simulated} Here, quantum tunneling could provide a large  advantage,\cite{boixo2016computational} in particular when the energy landscapes display tall but thin barriers,  which  are easier to tunnel through quantum-mechanically, rather than to climb over by means of thermally activated rare events, whose frequency is exponentially suppressed as the height of the barrier increases.

In general, simulating real-time quantum dynamics requires the direct integration of the time dependent Schr\"{o}dinger equation. This is a formidable task as the Hilbert space of the systems grows exponentially with the number of constituents, which makes the unitary evolution of a quantum system only possible for fairly small problem sizes, on the order of  $40$ to 50 spins.
The characterization of quantum dynamics simplifies when it is dominated by tunneling events. In this case,  the useful quantities we want to predict is the transition rate between initial and final state (e.g. reactants and product in chemical reactions), and the pathway of the transition.

\begin{figure}[b]
 \includegraphics[width=0.9\columnwidth]{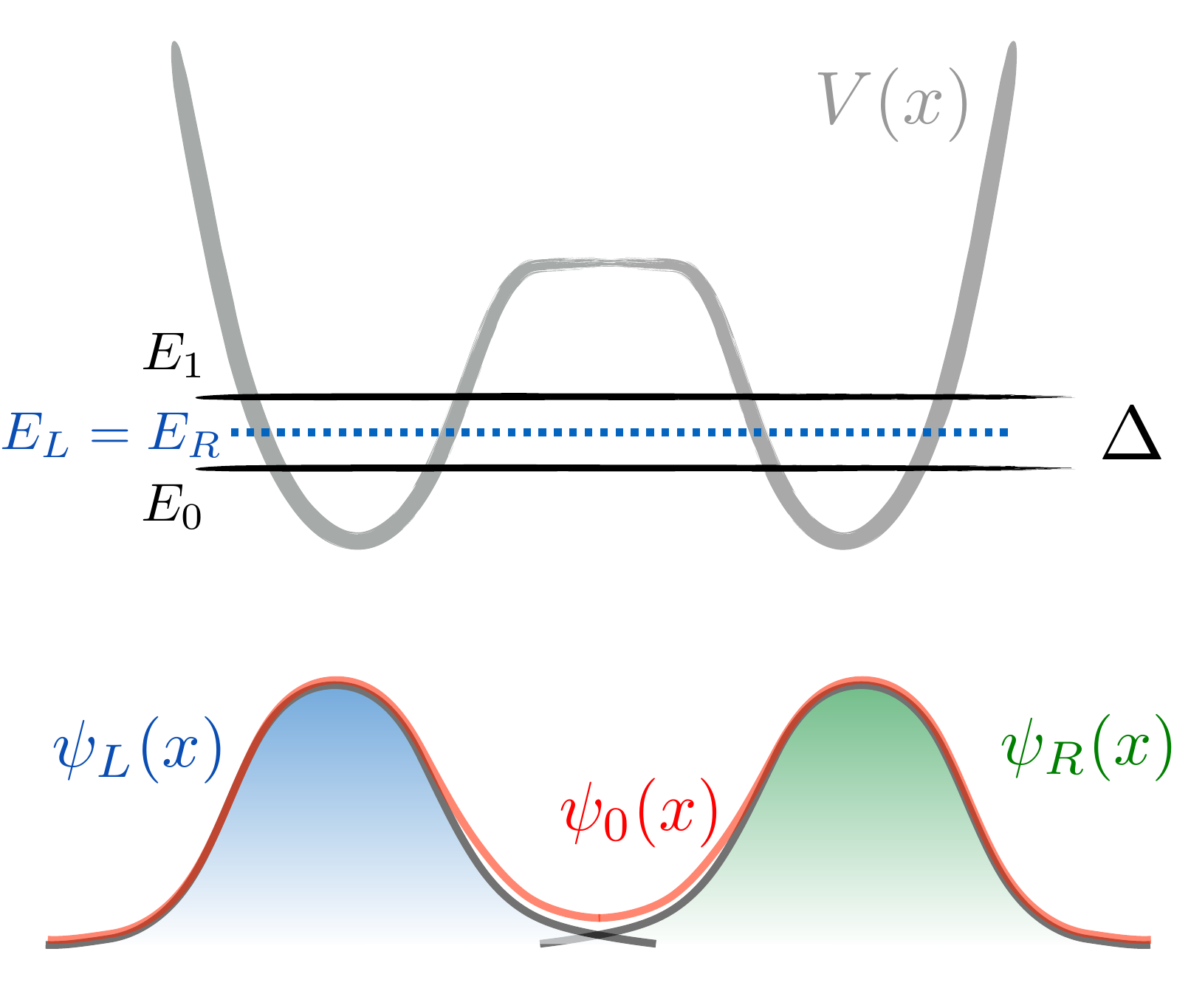}
 \caption{ (color online).  Cartoon of a double well potential energy $V(x)$ and energy levels. The degenerate levels $E_L$ and $E_R$ correspond to the localized states $|\psi_L\rangle$ (blue) and $|\psi_R\rangle$ (green). The degeneracy is lifted by the linear combination of localized states with produce the true eigenstates $\psi_0 = 1/\sqrt 2( |\psi_L\rangle +  |\psi_R\rangle)$ (red curve) and $\psi_1 = 1/\sqrt 2( |\psi_L\rangle -  |\psi_R\rangle)$.
 The tunneling splitting $\Delta$ can be calculated from the overlap of the localized states.
 }
 \label{fig:doppia}
 \end{figure}

For simplicity let us first consider tunneling in a deep double-well system, well described by the lowest two eigenstates  of the unperturbed tunneling system.$|\psi_0\rangle$,  $|\psi_1\rangle$ which can be expressed as linear combinations of the degenerate states $|\psi_L\rangle$ and $|\psi_R\rangle$, localized respectively in the left and right well (see Fig.~\ref{fig:doppia}).
The isolated system exhibits characteristic oscillatory behavior between the unperturbed states, $| \psi_L \rangle$ and $| \psi_R \rangle$, under the action of the Hamiltonian $H$, with frequency proportional to the tunneling matrix element $\langle \psi_L | H |  \psi_R \rangle =\Delta/2$.

Coherence is easily destroyed by the presence of external noise, as is the case in the proton transfer reactions and in QA. Coupling to an environment can then stop the oscillatory and the transition rate is given by the \emph{incoherent} tunneling rate, proportional to $\Delta^2$.\cite{weiss1987incoherent} 
This is also the relevant tunneling rate in the adiabatic evolution of quantum annealing (QA), where the annealing time must scale as $\Delta^{-2}$, in order to avoid Landau-Zener diabatic transitions from the ground state to the first excited state.\cite{farhi2000quantum,PhysRevE.58.5355}

QMT also appears in QMC simulations, which can be efficient and scale polynomially with system size for quantum many-body problems without a sign problem ( i.e. that the system should obey bosonic statistics or distinguishable particles). Path integral Monte Carlo (PIMC) has been successfully applied to a broad range of continuum  and lattice models. In particular PIMC simulations\cite{berne1986simulation,Ceperley:1995p19543}  have addressed problems in which QMT is important, such as proton delocalization in water\cite{ceriotti2013nuclear,carwater}, hydrogen\cite{chen2013quantum} and QA.\cite{Santoro29032002,Heim12032015}

PIMC is based on the path integral formalism of quantum mechanics and samples the density matrix corresponding to the quantum Hamiltonian $H$ by means of a classical Hamiltonian $H_{cl}$ on an extended system having an additional dimension, the \emph{imaginary time} direction.
The original quantum system is thus mapped into a classical one, which can be simulated by standard Monte Carlo sampling.
 
 Although QMC techniques are rigorously derived to describe equilibrium properties, we here show that equilibrium PIMC simulations also provide important dynamical quantities, and in particular the the quantum tunneling rate. In Ref.~\onlinecite{isakov2015understanding} we have studied tunneling events in a ferromagnetic Ising model. The Ising ferromagnet can be described by an effective double well model, with the total scalar magnetization as  reaction coordinate. We have numerically demonstrated that PIMC tunneling events occur  with a rate $k$ which scales, to leading exponential order, as $\Delta^2$ -- identical to the physical dynamics. We have also seen that with open boundary conditions (OBC) in imaginary time, the tunneling rate can becomes $\Delta$, thus providing a quadratic speed-up.

In this paper we investigate the scaling relation between the PIMC tunneling rate and $\Delta$ for a broader class of problems, of paradigmatic importance in quantum chemistry. We explore  models where the effective one-dimensional picture of tunneling should break down.\cite{japan} Our results for continous variables extend the ones for the Ising model\cite{isakov2015understanding} and we find that  the QMC tunneling rate always follow the $\Delta^2$ scaling (or better with OBC). We argue that this is a manifestation of a  general phenomenon, that QMC can efficiently simulate the tunneling splitting of the ground state energy levels in  multidimensional systems.

\section{Instantons and QMC}

\subsection{Path Integral Monte Carlo}

PIMC and path integral molecular dynamic (PIMD)  techniques  directly arise from the Feynman path integral formulation of quantum mechanics and are used to simulate thermodynamic equilibrium.
To briefly introduce this approach for continuous space we start from the expression for the partition function $Z$:
\begin{equation}
\label{eq:Z}
Z = \int dx \langle x | e^{-\beta H} | x \rangle
\end{equation}
where $x$ is the particle position (the generalization to arbitrary dimensions is straightforward), $\beta=1/k_BT$ is the inverse temperature and $H$ is the Hamiltonian of the system. 
Typical real space Hamiltonians are sums of two non-commuting operators $H=\Theta + V$, where $\Theta=1/2m ~ \partial^2 / \partial x^2$ is the kinetic operator ($m$ being the particle mass), and $V(x)$ is the potential energy.
We first notice that the operator $e^{-\beta H}$ corresponds to an evolution in imaginary time $\beta$. We use the Trotter-Suzuki approximation  $e^{-\delta_\tau (\Theta+V)} \approx e^{-\delta_\tau \Theta }e^{-\delta_\tau V}$ for small $\delta_\tau$.\cite{Ceperley:1995p19543}

Splitting the imaginary time evolution into $P$ small time steps of length $\delta_\tau = \beta/P$, the path integral expression for Eq.~(\ref{eq:Z}) then becomes
 \begin{equation}
 \label{e:pi}
 Z \propto \int d x_1 d x_2 \cdots d x_P \exp{\sum_{i=1}^P S_i }~,
 \end{equation}
where $S_i = K_i + U_i$ is the \emph{action} of each step.
$K_i= (x_{i-1}-x_i)^2/(2\delta_\tau/m)$ is the kinetic part and $U_i= \delta_\tau/2(V(x_{i-1} - x_i)$, in the so-called \emph{primitive} approximation.
Notice that $x_1 = x_P$ (closed boundary conditions in imaginary time), for evaluating the trace of the density operator.   

This provides an analogy between a quantum system and a classical system with an additional dimension: Eq.~(\ref{e:pi}) is a classical configurational integral and the multidimensional object $(x_1,\cdots,x_{P-1})\equiv \bx(\tau)$ can be viewed as a \emph{ring-polymer}, whose elements are connected by springs. Each element is labeled by its position along the imaginary time axis, with $0 \leq \tau < \beta$. 
We refer to the Ref.~\onlinecite{RevModPhys.67.279} for a detailed review of path-integrals.
An essential feature of Eq.~(\ref{e:pi}) is that the integrand is positive, and hence the distribution $\exp{\sum_{i=1}^P S_i }$ can be sampled by means of Metropolis Monte Carlo methods or Molecular Dynamics (MD) simulations. 
The main difference between a pure Monte Carlo vs a MD approach is that the latter samples from the canonical distribution by evolving an appropriate equation of motion, whereas the former uses stochastic  Monte Carlo dynamics.

\subsection{Instantons in PIMC}
\label{s:insta}

Connections between exact quantum dynamics and PIMD approaches, such as Centroid Molecular Dynamics\cite{cmd} and Ring Polymer Molecular Dynamics\cite{rpmd} have been discussed \cite{voth,shorttime,hele} in the context of real space simulations. Here we follow an alternative approach and summarize the picture of Refs.~\onlinecite{isakov2015understanding,jiang2016scaling} based on the \emph{instanton} theory of tunneling through energy barriers.

   \begin{figure}[t]
 \includegraphics[width=1\columnwidth]{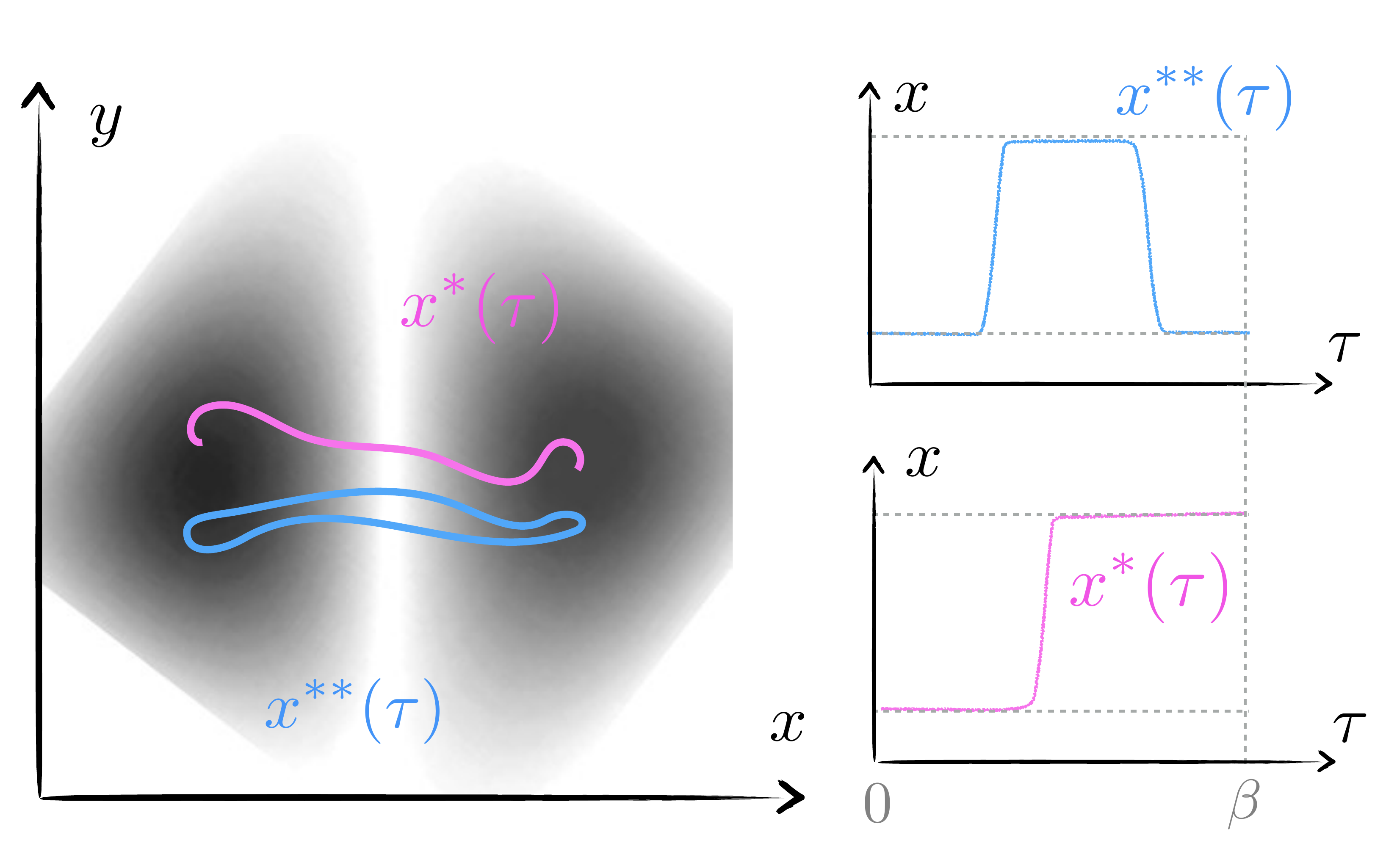}
 \caption{ (color online). \emph{Left.} Cartoon of the typical instantonic paths in configuration space, with PBC, $\bx^{**}(\tau)$ (cyan) and OBC in imaginary time, $\bx^{*}(\tau)$ (pink). These paths are transition states of the PIMC and PIGS pseudodynamics respectively (in the space of imaginary time trajectories) in double well models (sketched in the grey scale heatmap, see Fig.~\ref{fig:malo} for a more realistic example). \emph{Right.} Instantonic trajectories (projected ont ethe reaction cooordinate $x$ axis) as a function of the imaginary time $\tau$. Notice that PIMC instantons have to cross twice the barrier to fulfill the PBC constrain.
 }
 \label{fig:cart}
 \end{figure}

In a PIMC or PIMD simulation one samples paths $\bx(\tau,t)$ at each update along the simulation time axis $t$, and these paths are distributed according the  functional $ S(\bx(\tau)) $ as in Eq.~(\ref{e:pi}).
We can define an underlying pseudodynamics used to sample the paths to be given by a first order Langevin dynamics, 
$\partial \bx(\tau,t) / \partial t = - \delta S / \delta \bx(\tau,t) + \eta(\tau,t)$.
In this case the analogy between quantum statistic and classical statistical mechanics has  already been worked out in the \emph{stochastic quantization} approach   in the context of quantum field theory.\cite{parisi1981perturbation}
Here, the velocity of the (deformations of) path $\partial \bx(\tau,t) / \partial t$ is linked to the generalized force $\delta S / \delta \bx(\tau,t)$ and a Gaussian white noise $\eta(\tau,t)$ satisfying the obvious fluctuation-dissipation relation.
We can numerically integrate  the discretized version of the equation of motion (with time-step $\delta_t$), 
$\bx(\tau,t+\delta_t) = \bx(\tau, t)  -  \delta _t ~\delta S / \delta \bx(\tau,t) +\sqrt{2  \delta_t} {\bf z} (\tau,t)$, where $ {\bf z} (\tau,t)$ is a deformation path, which, after a Trotter discretization is a vector of uniformly random distributed number in the range $[-1,1]$.
This defines a Markov chain whose fixed point is the desired distribution, in the $\delta_t \rightarrow 0$ limit.

If the system displays two degenerate minima, then the transition state of the pseudodynamics is given by the point $\bx_{TS}(\tau)$ satisfying $\delta S( \bx_{TS}(\tau))/ \delta \bx(\tau) = 0$ with the condition that  $\bx_{TS}(\tau)$  is not entirely contained  in one of the attraction basins corresponding to the two minima\cite{parisi1981perturbation,sega2007quantitative,autieri2009dominant,mazzola2011fluctuations}.

Finding this transition state is generally very complicated, but in the case of a double well potential $V(x)$ it can be done analytically.
Here the dominant contribution to the integral comes from the stationary action path $\bx^{**}(\tau)$ (determined exactly by the condition
 $\delta S( \bx(\tau))/ \delta \bx(\tau) = 0$) which is called instanton\cite{coleman_fate_1977,forkel,Chudnovsky-book}.
 This trajectory in imaginary time corresponds to a particle moving in the inverted potential $-V(x)$ (see Fig.~\ref{fig:cart}).
Following Ref.~\onlinecite{isakov2015understanding} it is possible to evaluate the action $S$ at this point and
the amplitude is given by
\begin{equation}
  \exp({-S[\bx^*(\tau)]})  \propto \Delta \quad \textrm(instanton),
\end{equation} 
where $\bx^*(\tau)$ is  the open trajectory which connects the two classical turning points under the barrier, near the minima.
Notice that, when computing the (diagonal) density matrix $\rho(x)$ periodic boundary conditions (PBC) in imaginary time are required.
Now the integral over the \emph{closed} paths it is dominated by the imaginary time trajectory $\bx^{**}(\tau)$ that moves under the barrier starting, reaches the turning point, and returns.
Therefore, the saddle point estimation of the integral  gives a {\it squared} tunneling amplitude
\begin{equation}
  \exp({-S[\bx^{**}(\tau)]}) \propto \Delta^2  \quad \textrm(double~ instanton),
\end{equation}
due to the cost of creating an instanton and an anti-instanton (see Fig.~\ref{fig:cart}).
Returning to the PIMD pseudodynamics, 
according to Kramers theory\cite{hanggi1990reaction}, the escape rate is $k \propto e^{- S(\bx_{TS})}$, and therefore $k \propto \Delta^2$ if standard closed path integrals are used, whereas  $k \propto \Delta$ if the paths are opened.
In Sect.~\ref{s:dw} we extend the study of Ref.~\onlinecite{isakov2015understanding} and demonstrate that the quadratic speedup in the tunneling rate in the case of open boundary path integrals holds also in multi-dimensional continuous space problems.

\section{One-dimensional double well potential}
\label{s:dw}
\begin{figure}[t]
 \includegraphics[width=1\columnwidth]{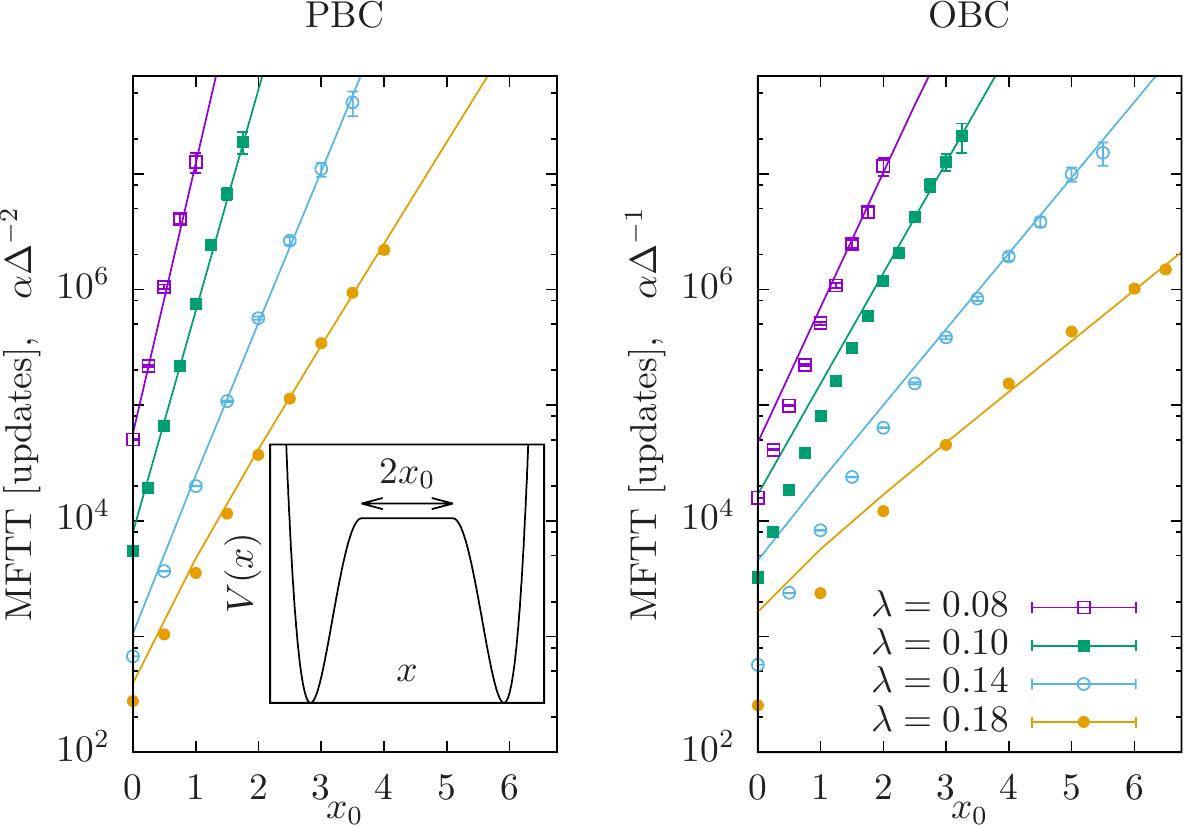}
 \caption{ (color online). Average  MFTT tunneling time with PIMC (for PBC and OBC) as a function of $x_0$ for different values of $\lambda$, at $\beta=20$, corresponding to a temperature always much lower than the barrier height. We use a dimensionless mass parameter $m=1/2$.
  The inset shows the shape of double well potential $V(x)$, which barrier width (at top) is $2 x_0$. 
   Notice that for OBC, the measured MFTT is smaller than the one predicted by the $1/\Delta$ formula, when the tunneling rate is large. This happens because both the 
 $\bx^{*}(\tau)$ and the $\bx^{**}(\tau)$ channel contribute to the tunneling.
 }
 \label{fig:gaps}
 \end{figure}

Let us consider the following one-dimensional double well potential,
\begin{align}
\label{e:pot}
V(x) =\begin{cases} 
      \lambda (x-x_0)^4 - (x-x_0)^2, & x\geq x_0 \\
      0, & -x_0\le  x\le x_0 \\
       \lambda (x+x_0)^4 - (x+x_0)^2, & x\leq -x_0
   \end{cases}
\end{align}
with $\lambda,x_0 > 0$.
 We can separately tune  the width and the height of the barrier, varying $\lambda$ and $x_0$. The height of energy barrier is $\Delta V = 1/4 \lambda$, and the distance between the two minima is 
$d  = 2 (x_0 + \sqrt{1/ 2\lambda})$ (see inset of Fig.~\ref{fig:gaps})
Decreasing $\lambda$ reduces the energy splitting $\Delta$, as the two wells become deeper and more separated.
The parameter $x_0$  only increases the well separation but doesn't change the potential energy barrier height.
Moreover, a variation of $x_0$ leaves  the characteristic frequency of the potential wells unchanged,  i.e. the kinetic  energy associated to the localized states $| \psi_L \rangle$ and $| \psi_R \rangle$. 

 Following Ref.~\onlinecite{isakov2015understanding} we measure the mean first tunneling time (MFTT), defined as the number of updates required to find the system in the right well, if  the particle has been localized in the left one at the beginning of the simulation.
 From Fig.~\ref{fig:gaps} we see that the MFTT scales as $1/\Delta^2$ when PBC are used, whereas it scales as  $1/\Delta$ for OBC, as the parameters $x_0$ and $\lambda$ change.
  The gap $\Delta$ is obtained using a discrete variable representation (DVR) technique\cite{colbert1992novel}.
  This scaling relation holds for PIMC with local Metropolis updates and PIMD  (using both first and second order Langevin thermostats), at large $\beta$, and in the limit of small time steps $\delta_\tau,\delta_t \textrm{(for PIMD)} \rightarrow 0$ limit.
This means that the scaling of tunneling rate in a double well model $k \propto \Delta^2$ is correctly reproduced.\cite{weiss1987incoherent}.

   \begin{figure}[t]
 \includegraphics[width=1\columnwidth]{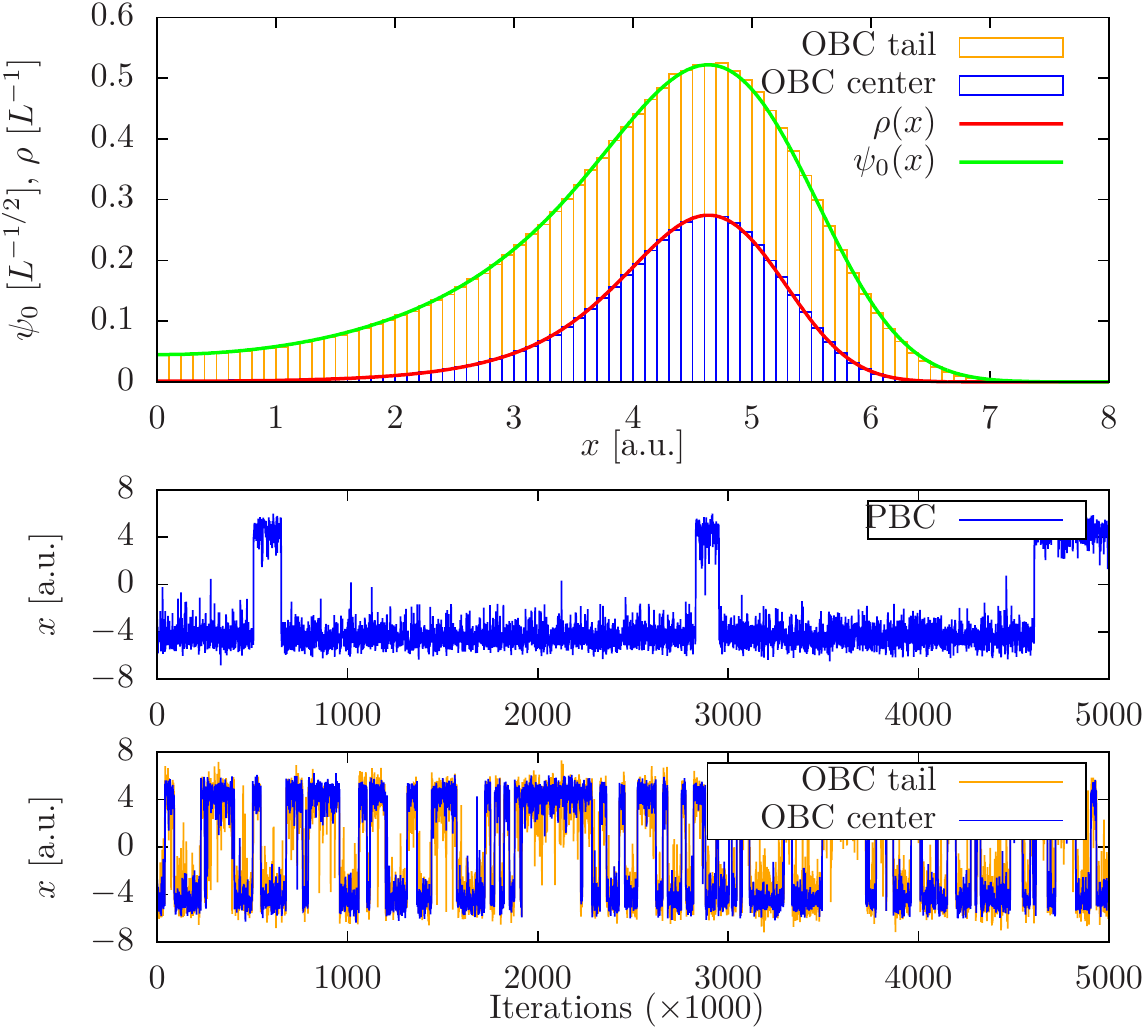}
 \caption{ (color online). Top panel: position distributions (histograms) obtained considering the center (blue) or the tail (orange) of the OBC path. The distributions are area-normalized respectively with the exact $\rho(x)\approx |\psi_0|^2$ distribution (red) and the exact ground state $\psi_0(x)$ (green). 
We plot only for $x>0$ and we use $x_0=3$ and $\lambda = 0.14$ in Eq.~(\ref{e:pot}). 
 The difference between the sampled distributions and the reference ones are negligible. We perform simulations at low temperatures, $\beta=20 \gg \Delta V$.
 Middle and lower panel: the position of the particle as the simulation progresses for PBC and OBC (both for center and tail). As expected the tunneling rate is much larger for OBC.
 }
 \label{fig:dw}
 \end{figure}

Why do the open paths tunnel faster from the point of view of PIMC pseudodynamics?
To answer this question  we first observe 
 that, for sufficiently low temperatures,  the center of the open-path $\bx^{*}(\tau \approx\beta/2)$  sample  from the ground state distribution $|\psi_0(x)|^2$, whereas the tails, $\bx^{*}(\tau \approx0)$ and  $\bx^{*}(\tau \approx\beta)$, sample from the ground state distribution $\psi_0(x)$. 
Therefore the tails spend more time inside the barrier  (see Fig.~\ref{fig:dw}) compared  to the center,  which follows instead the more localized $\psi_0^2$ distribution.
Once that one of the two tails crosses the barrier, then the rest of the open-polymer may easily follow, so that the whole polymer "tunnel" faster compare to its PBC counterpart.
This also means that, with  OBC,  
 it is possible to sample from the equilibrium distribution  $\rho(x)\approx  |\psi_0|^2$, using the center of the path, while having a considerable speed-up in the sampling.
We notice that this feature is not surprising as the OBC technique is closely related to the so called \emph{path integral ground state}\cite{pigs} (PIGS) technique.
Indeed, in the PIGS\cite{pigs} approach, sampling from the tails gives  the mixed distribution $\psi_0(x) \psi_T(x)$, but in our case the trial wavefunction is $\psi_T(x)=1$.
Therefore, we propose that OBC should be used not only in the context of  quantum annealing but much more broadly also in material simulations,
as far as low-temperature conditions are investigated.

\section{Multidimensional tunneling}

The double well model provided in Sect.~\ref{s:dw} is a prototypical example of one-dimensional tunneling.
One could argue that, despite having many-spins degrees of freedom, the spin models investigated in Ref.~\onlinecite{isakov2015understanding}  are also effectively one-dimensional models, as the relevant  reaction coordinate is the total scalar magnetization $M$.
Indeed the \emph{instantonic} nature of the transition state can be seen if we plot  $M(\tau)$ as a function of the imaginary time parameter $\tau$.

It is much more straightforward to devise models that require multidimensional tunneling in continuos space, rather than spin models.\cite{PhysRevB.73.144302,inack2015simulated}
To this end we borrow insights from  quantum chemistry, where simplified model for characterizing proton tunneling have been devised.\cite{auerbach1985path,makri1987,japan,nakamura2013quantum}
In particular in Ref.~\onlinecite{japan} a semiclassical theory of multidimensional tunneling is formulated, unraveling its qualitative differences compared to  one-dimensional tunneling.
It was found that in multidimensional tunneling two regimes can be identified: the \emph{pure tunneling} case, which is effectively one-dimensional, where
the tunneling path can be defined uniquely, and the \emph{mixed tunneling} regime when tunneling occurs very broadly, i.e. where a set of dominant semiclassical paths $\{ {\bf x}_{TS} \}$ is not defined.
In the first case, the action which defines the semiclassical wavefunction is purely imaginary, whereas in the latter the action is complex.
We refer the interested reader to Ref.~\onlinecite{japan} for the analytical details.

Investigating QMC simulations for such mixed tunneling models, where the QMC scaling relation with the exact QMT rate might be expected to break down, we instead find that  the  quantum tunneling rate given by QMC scales as  the adiabatic quantum evolution also in this case.

\subsection{QMC tunneling rate scaling}

We first consider the simple \emph{shifted parabola} bidimensional model of Ref.~\onlinecite{japan}, which is a minimal model for the antisymmetric mode coupling mechanism for proton tunneling in malonaldehyde, a well-studied molecular test case.
The Hamiltonian reads
\begin{equation}
H = \Theta + V_{A}~,
\end{equation}
with 
\begin{equation}
 \Theta = - { g^2 \over 2} \left( {\partial^2 \over \partial x^2} + {\partial^2 \over \partial y^2} \right)~,
\end{equation}
where $g>0$ is a dimensionless parameter which sets the strength of the quantum fluctuations.
The potential is 
\begin{align}
\label{e:potsp}
V_{A}(x,y) =\begin{cases} 
     \frac 12 (x+1)^2 + \frac 12 \omega_y^2 (y+y_0)^2, & x < 0 \\
   \frac 12 (x -1)^2 + \frac 12 \omega_y^2 (y-y_0)^2, & x\geq 0~,
   \end{cases}
\end{align}
where $y_0>0$ and $\omega_y^2>0$ are dimensionless harmonic potential parameters.
This potential represents two parabolas, located respectively in the half-planes $x<0$ and $ x>0$, with centers shifted along the $y$-axis by an amount $2 y_0$.
In the case of malonaldehyde molecule, the coordinate
$x$  represents the motion of transferring the proton, while $y$ represents the C--O stretching mode.
The sign of $a^2=y_0^2-g/\omega_y$, distinguishes between the two ground state QMT cases: pure tunneling  for $a^2> 0$ and mixed tunneling for $a^2<0$.

   \begin{figure}[t]
 \includegraphics[width=1\columnwidth]{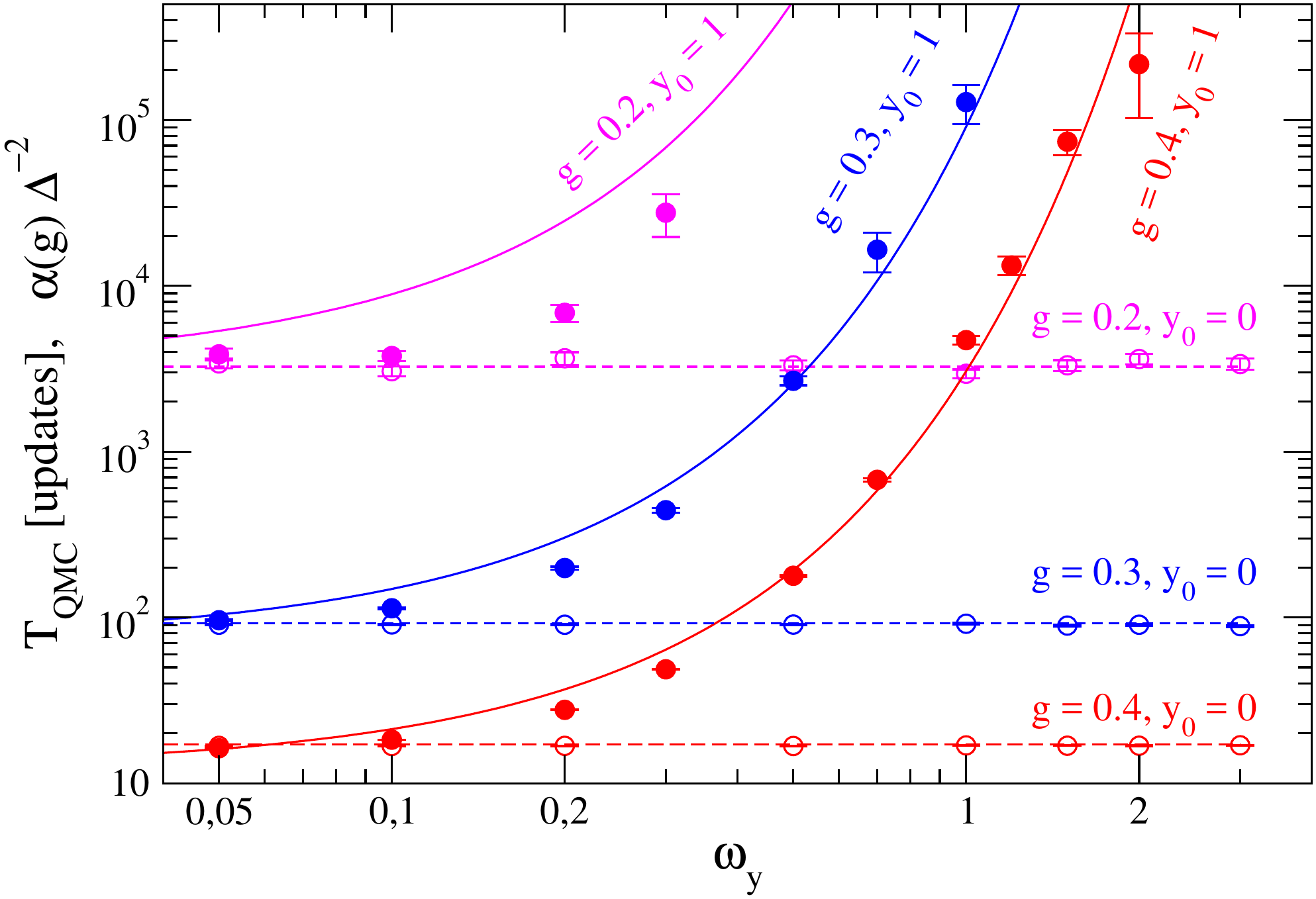}
 \caption{ (color online). Average  MFTT tunneling time with PIMC (with PBC) as a function of $\omega_y$ for different values of $g=0.2,0.3,0.4$ (pink, blue and red data series respectively) and two values of $y_0=0$ (empty symbols) and $1$ (full symbols). The potential considered is $V_A$ as in Eq.~\ref{e:potsp}.
 Lines represent fit to the exact $\Delta^{-2}$ gap values, obtained with the DVR method. The proportionality constant $\alpha(g)$ which multiplies the inverse gap squared is different for each $g$ value, and fitted using only the $y_0=0$ data series.
 Notice the  logarithmic scale on both axis.
 }
 \label{fig:2d}
 \end{figure}

In Fig.~\ref{fig:2d} we present results of PIMC simulations with local updates, using PBC, at large $\beta$ (very low temperature),
and in the converged time step $\delta_\tau \rightarrow 0$ limit, to describe faithfully ground state tunneling.
The path deformations are obtained by displacing each bead at a time by an amout $(dx,dy)$. The displacements are Gaussian distributed with zero mean and the variance is tuned in order to obtain a Metropolis acceptance probability of $\approx 40\%$. 

Again we study the MFTT obtained with PIMC simulations as a function of the parameter $\omega_y$, in the range $[0.05,2]$ and for three different choices of $g=0.2,0.3,0.4$, and for two shifting values $y_0=0$ and $1$.
Following Ref.~\onlinecite{isakov2015understanding} we define the MFTT as the number of PIMC updates required to observe an instantonic state. In turn, we algorithmically define an instanton path, as spending approximately the same fraction of imaginary time in either well.

With these parameters ranges\footnote{The actual value would be $g\approx0.1$, here we artificially enhance $\hbar$ in order to increase the observed tunneling rate. }  we can roughly mimic  proton transfer reactions in malonaldehyde.\cite{makri1987}
For this molecule, it is found that, if the tunneling is described only by a one-dimensional process, the tunneling rate is reduced by  two orders of magnitude compared to experimental and recent theoretical values\cite{moreno1990,malo2016}. Furthermore, it was argued in Ref.~\onlinecite{japan} that deviations from the one dimensional picture leads to a mixed tunneling regime where no well defined tunneling path exists.
Therefore it could be possible that QMC underestimates the exact tunneling rate. In the context of quantum annealing problem,  this might mean that the performances of QA and its simulated version through QMC could be very different, under these ``mixed tunneling"  conditions.

We first perform tests for $y_0=0$, where we are always in the regime $a^2 < 0$ , i.e. the mixed tunneling regime. The gap $\Delta$ is constant as a function of $\omega_y$, and we observe the same in QMC, where the MFTT remains constant. Its precise value depends on the parameter $g$.
We use these data to fix the proportionality constant $\alpha(g)$, which we use later to compare the MFTT to the value $\alpha(g)\Delta^{-2}$.
Notice that in the $\omega_y \rightarrow 0$ limit the two wells become parallel and indefinitely extended along the $y$ direction. 
In this limit, we observe an \emph{infinite} number of tunneling paths that connects reactants on left well with the products on the right.

Next, we set $y_0=1$ and repeat the simulations. This time the scaling of $\Delta^{-2}$ as a function of $\omega_y$ is non-trivial.
Nevertheless it approaches the previous value -- for any given $g$ -- in the $\omega_y \rightarrow 0$ limit, as the two wells are infinitely long and the shift given by $y_0$ becomes irrelevant.
In this case, we cross the transition point $a^2=0$ between the the two regimes of tunneling, when $\omega_y = g$.
We still observe a satisfactorily agreement between the QMC MFTT data series and the $\alpha(g)\Delta^{-2}$ functions.
While a residual difference  between the QMC data and the expected $\Delta^{-2}$ behaviour can now be appreciated, this difference is small over the  broad $\omega_y$-range investigated, i.e. nowhere near the 2 order of magnitude worst case scenario, reported in Ref.~\onlinecite{moreno1990}.
We note that in Ref.~\onlinecite{moreno1990}  the full multidimensional problem is reduced to a one-dimensional calculation. This is the origin of the observed large deviation from the theoretical value.
Notice also that, once we fix the constant $\alpha(g)$ the QMC tunneling time is always smaller than $\alpha(g)\Delta^{-2}$, so QMC tunnels slightly more efficiently than QA, even with PBC.

\subsection{QMC reaction pathways and fluctuations around the instanton solution}

In this Section we explicitly track  the QMC pseudo-dynamics transition states and  compare   to the  instantonic trajectory computed by minimizing the action $S$.
Let us consider the \emph{symmetric} mode coupling potential,
\begin{equation}
\label{e:potsym}
V_{S}(x,y) = \frac 1 8 (x-1)^2(x+1)^2 + {\omega_y^2 \over 2} \left( y + {\gamma \over \omega_y^2} (x^2+1)   \right)^2~.
\end{equation}
This potential energy surface is continuos and has been widely used as a model for proton tunneling. In the typical example  of malonaldehyde,  the coordinate $x$  represents the motion of the proton transferring between the Oxigen atoms, while $y$ gives the scissors-like motion of the O-C-C-C-O frame\cite{japan}.
We use the dimensionless potential parameters $(\omega_y,\gamma,g) = (0.48,0.39,0.10)$ to fit the model to the \emph{ab-initio} potential energy surface\cite{moreno1990}.
In this way we can directly compare the transition paths given by the PIMC simulation with other techniques, such as the \emph{Ring Polymer Instanton} (RPI) method,\cite{richardson2011ring,richardson2011instanton,malo2016,richardson2016concerted} recently introduced to calculate energy splitting.
In the RPI framework, one first needs to locate the saddle point of the action $S$ (the instanton), and then evaluates the splitting energy by computing the functional integral up to second order in the fluctuations around the dominant contribution.
This approach employs neither PIMC sampling nor PIMD, as the instanton path is obtained via action's minimization and the initial guess is an OBC path  that already connects the reactant to product state.

\begin{figure}[ht]
 \includegraphics[width=1\columnwidth]{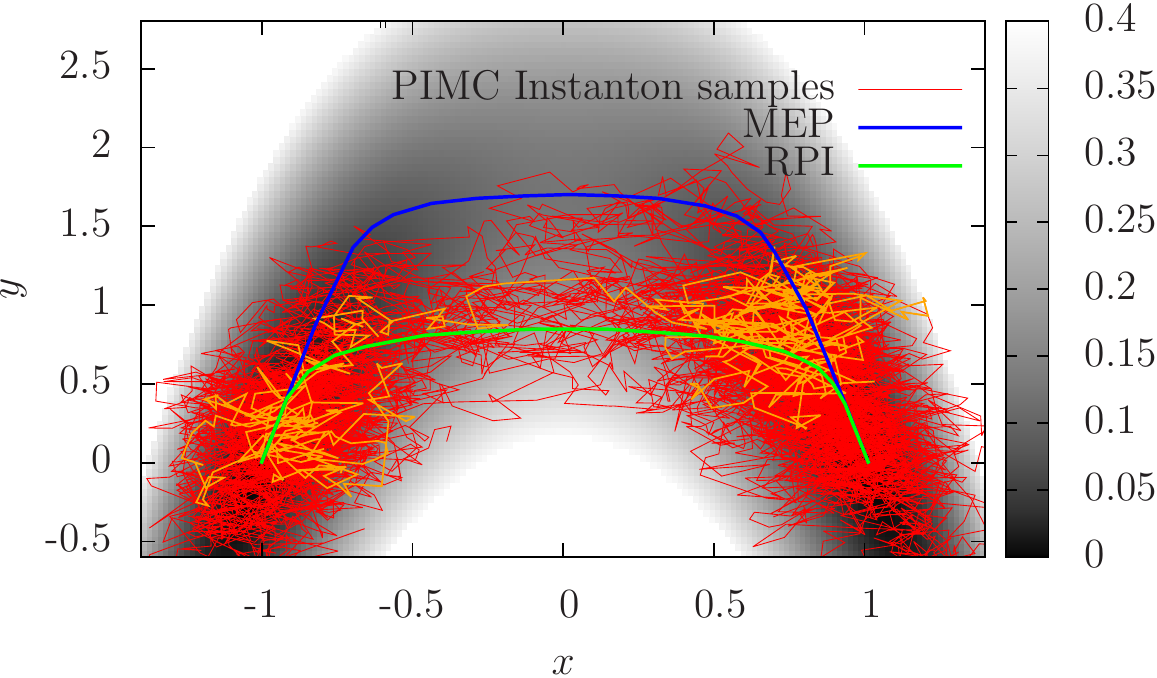}
 \caption{ (color online). Heat map plot of the two-dimensional model potential $V_S(x,y)$ of malonaldehyde of Eq.~\ref{e:potsym}.  Red lines represents a collection of $20$ instanton paths sampled with PIMC. 
 These samples are uncorrelated as they correspond to independent tunneling events after full reinitialization of the starting path in the reactant well, i.e. when the MFTT identification criteria are met (see text) we stop the PIMC simulation and collect the last path which has been generated.
 We use a sufficiently large inverse temperature $\beta=400$, $P=512$ Trotter slices and OBC in imaginary time. We single out, in orange, one of these instances, in order to appreciate their \emph{instantonic} character. Indeed most of the Trotter slices are located at the bottom of the two wells, and only very few of them on the barrier (cfn. Fig.~\ref{fig:cart}), these correspond to the middle of the imaginary time trajectory ($\tau \approx \beta/2$). 
 The PIMC instanton paths are not smooth, given the large number of Trotter slices, and represent themselves fluctuations around an average transition path, qualitatively very close to the RPI solution taken from Ref.~\onlinecite{malo2016} (green, see text). We also plot the MEP (blue) for comparison. The proton tunneling paths typically take place on a region quite far from the saddle point $(0, \gamma/\omega_y^2 \approx 1.7)$ of the potential.
 }
 \label{fig:malo}
 \end{figure}

In Fig.~\ref{fig:malo} we plot a sample of transition paths  produced by the PIMC pseudodynamics and  recognize their instantonic character.
We compare some OBC  transition paths sampled with our PIMC simulation, against the RPI solution recently published in Ref.~\onlinecite{malo2016}.
We see that these instanton paths form a bundle around the RPI saddle point solution, and are qualitatively distant from the minimum energy path (MEP), which  would be typical of a classical thermally activated process\cite{malo2016}.
It is remarkable that a simple PIMC simulation obtains the instanton path, which is otherwise computed only by a complex minimization procedure as in the RPI scheme. 

Another advantage of PIMC is that we can directly sample the statistical fluctuations around the dominant solution $\bx^*(\tau)$.
To second order the action can be expanded as  \cite{parisi1981perturbation}
\begin{align}
\label{seff3}
S & \simeq S[\bx^*] + \frac{1}{2} \int_0^\beta d \tau' \int_0^\beta d \tau'' 
\frac{\delta^{2} S[\bx^*]}{\delta \bx(\tau')\delta \bx(\tau'') } \by(\tau)\by(\tau) \nonumber \\
	& \simeq S[\bx^*] +  \frac{1}{2} \int_0^\beta d \tau ~~ \by(\tau) \hat{G}[\bx^*] \by(\tau),
\end{align}
where $\by(\tau)$ is a fluctuation path, satisfying $\by(0)=\by(\beta)=\mathbf{0}$, and  
$\hat{G} = -\frac{d^{2}}{d\tau^{2}}+V''[\bx^*]$ is 
the fluctuation operator, or Hessian, adopting the notation of Ref.~\onlinecite{richardson2011ring}.
Here $V''[\bx^*]$ is the second derivative of the potential computed along the instanton path $\bx^*(\tau)$.  

In practice one always deals with discretized trajectories in imaginary time. Therefore also the operator $\hat{G}$ is discretized using finite differences and then diagonalized to obtain the normal modes and frequency of the fluctuations. The resulting product of Gaussian integrals allows one to evaluate Eq.~\ref{seff3}.

However, 
 it could be cumbersome to evaluate  $G$ for ab-initio potentials (as they require evaluation of second derivaties of the potential), or in the case of rugged energy landscapes, where local curvature at the saddle point $V''[\bx^*]$ does not correctly represent the actual amplitude of the quantum fluctuations\cite{pietro,mazzola2011fluctuations}.
On the other hand, the inverse operator $G^{-1}$ can be computed stochastically with PIMC, using the  relation
\begin{equation}
\hat{G}^{-1}[\bx^*](\tau_{1},\tau_{2})=\langle \by(\tau_{1})\by(\tau_{2})\rangle_{\bx^*},
\end{equation}
where the right hand side denotes the statistical average of the fluctuations, around a given path $\bx^*$.
This approach gives a more effective and fast estimate of the curvature of the potential surface in the above cases.

We note that PIMC sampling techniques have already been used to compute tunneling splittings in molecular and condensed matter systems\cite{kuki1987electron,ceperley1987calculation,alexandrou1988stochastic,matyus2016quantum}.
Here we propose an alternative and simple way to compute ratios of quantum mechanical rate constants. Suppose that the potential displays several minima, i.e. that we have one reactant state $R$ and two, or more, possible product states $P_1,P_2$.
By computing the ratio of the average PIMC tunneling times, with OBC, required to perform the transitions $R \rightarrow P_1$ and $R \rightarrow P_2$ respectively, we can estimate   the ratio of the tunneling splittings $\Delta_{R,P_2}/\Delta_{R,P_1}$ corresponding to the two quantum mechanical transitions, provided that enough statistics of instanton events can be gathered in a reasonable amount of time.
This approach is predictive, as the instantons are generated by the PIMC pseudodynamics, without any \emph{a-priori} knowledge of the final product state.

This problem is closely connected with quantum annealing, where, starting from some high energy ``reactant'' states $R$'s, one would like to reach the ``product'' state $P_i$ having the lowest possible potential energy, after a sequence of tunneling events\cite{knysh2015}.
The relative probability of finding this state, compared to other metastable ones, is well described by PIMC simulations.

\section{Conclusions}

We have studied the tunneling of path integral based equilibrium simulations in continuos space models, generalizing a previous study\cite{isakov2015understanding} on ferromagnetic spin models. We demonstrates that the PIMC tunneling rate scales as a $\Delta^{2}$ if periodic boundary conditions in imaginary time are used, while it scales as $\Delta$ with open boundary conditions.
These scaling relations seem to be a general property of path integral methods, as long as reasonable semi-local updates are employed during the Markov chain pseudodynamics (see Sect.~\ref{s:insta}).
In this case, in double well potentials, it is possible to directly identify the transition state of the path integral pseudodynamics --a purely classical process-- and therefore compute its classical reaction rate using Kramers theory.

This transition state is the \emph{instanton} path, and we remark here that this trajectory is sampled by the PIMC pseudodynamics using local updates, i.e. we don't need to engineer such kind of  global update moves as in  Refs.~\onlinecite{stella2006monte,alexandrou1988stochastic}.
Indeed the latter approach  would invalidate the premises and discussions presented in this paper and artificially increase the reaction rate observed with PIMC. On the other hand,  building in instantonic updates in the Metropolis procedure requires a full knowledge of the system, i.e. knowing in advance the transition states. Having this knowledge one would  solve beforehand the quantum annealing problem, for example, without even  running any PIMC simulation.

The quadratic speed-up in tunneling efficiency   is a robust feature of OBC simulations for tunneling through individual barriers.
In the context of simulations, therefore, we propose that  open path integral simulations should be used instead of PBC and will accelerate the sampling,  whenever ground state properties are desired.

We also turned our attention to  simplified models for proton transfer, where multidimensional tunneling is deemed to be important, and a semiclassical description of tunneling as an effective one-dimensional process has been seen to fail.
Nevertheless, the scaling relation of the PIMD transition rate, compared to the exact incoherent tunneling rate $\Delta^{2}$ holds also in this case.

The above finding is very interesting  because often in a multidimensional potential  the smooth tunneling path (instanton)  connecting the minima of the potential does not exists due to the effects of a so-called dynamical tunneling \cite{PhysRevA.41.32, japan,PhysRevA.72.024101,QuantumTheoryTunneling-book}. The lack of an instanton path was also studied,  e.g., in the case of the two-dimensional shifted parabola model Ref.~\onlinecite{japan} considered in our study. On the other hand, as explained  above in Sec. ~\ref{s:insta} based on the Kramers theory arguments, the existence of the instanton path  is a  key requirement that leads to the  identical QMT and PIMC scaling  laws.

To explain this conundrum we observe that the  difficulty with instanton description in the case of a tunneling in a multidimensional potential usually occurs when one needs to  match the  solutions given by Wentzel--Kramers--Brillouin (WKB) theory in classically allowed and forbidden regions at the boundary formed by caustics.  Caustics result in the complex  (oscillatory) behavior of the wavefunction under the barrier \cite{PhysRevA.41.32, japan,PhysRevA.72.024101}. This oscillatory behavior results in a phase problem in QMC. 

\begin{figure}[t]
 \includegraphics[width=1\columnwidth]{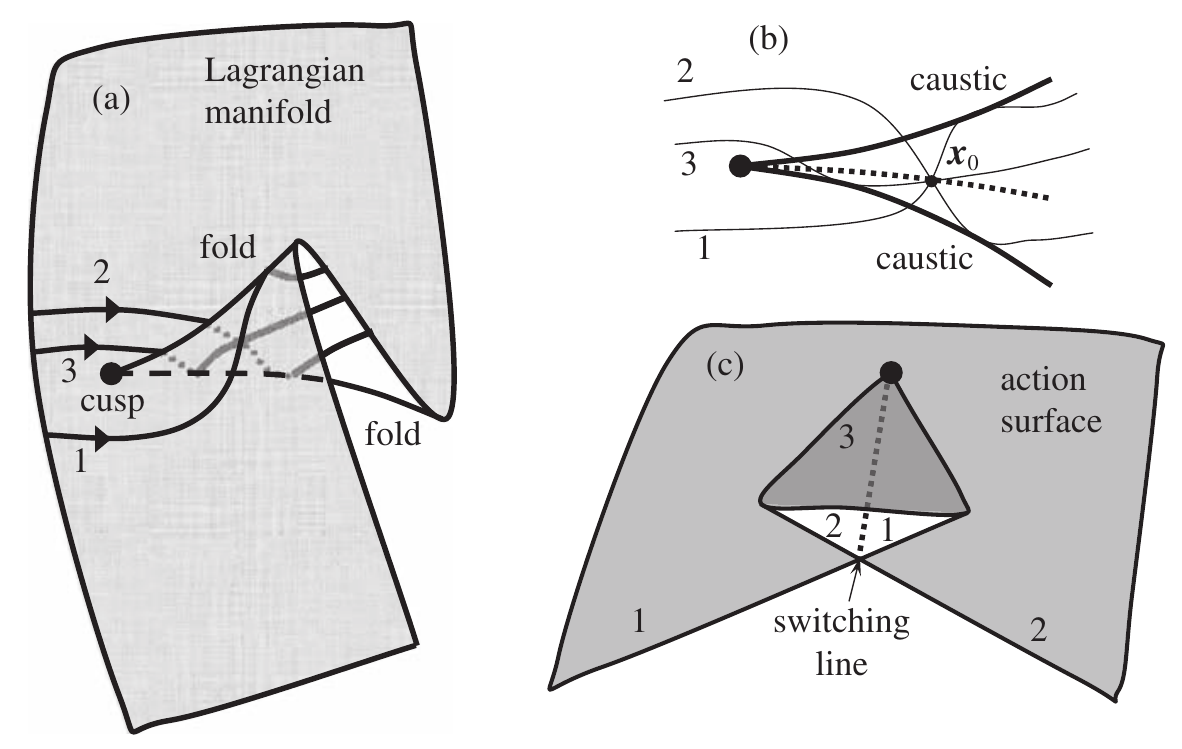}
 \caption{  (a) Unstable  manifold with a cusp singularity and three typical imaginary-time paths emanating from one of the minima of the potential. (b) Projection of the unstable  manifold
onto the coordinate plane ($X_1, X_2$) . The two folds project onto caustics, while projections
of the three trajectories intersect at the point $X_0$ that lies on the switching line  showed as   dashed. (c) The action $S(X_0)$ is a three valued
function in between the  caustics. Two of its lower branches intersect along the switching line. After  Refs.
~\onlinecite{kamenev2011field,dykman1994observable}.
 }
 \label{fig:sing}
 \end{figure}

We argue that at  zero temperature this situation does not occur generically, because a  classically allowed region in configuration space ``collapses" into the point corresponding to the minimum of the potential.   As usually, to  study tunneling one should consider   the wavefunction under the barrier that nearly  coincides with the ground state wavefunction near one of the minima of $V$,  exponentially decaying away from it. The mechanical action $S[{\bf x}(\tau)]=\int_{\infty}^{\tau}(m \dot {\bf x}^2(\tau_1)+V({\bf x}(\tau_1)))d\tau_1$ for the   wavefunction  is associated with the unstable Lagrangian manifold  \cite{guckenheimer2013nonlinear} formed by  real-valued trajectories $({\bf x}(\tau), {\bf p}(\tau))$ in the phase space  moving in the imaginary time $\tau$ in the   inverted potential $-V$ (above ${\bf p}(\tau)$ is a system momentum). The trajectories  emanate at time $t=-\infty$ from the corresponding maximum of the $-V$.    In general, projections of the Lagrangian manifold onto the coordinate space ${\bf x}$ can have caustics, cusps (and more complex singularities in the dimensions higher  then two\cite{Arnold:book}) in classically forbidden region. These singularities lead to  multi-valuedness of the action surface and some of its branches become complex. However the {\it minimum action} surface is real- and single-valued. It possesses lines  where the surface gradient is discontinuous  (see Fig.~\ref{fig:sing}) They correspond to the so-called switching lines  in configuration space \cite{dykman1994observable}. Points at different sides of the switching are reached by a topologically different  imaginary-time paths as shown in Fig.~\ref{fig:sing}(b). Therefore any point ${\bf x}$ can be reached by the most-probable path that provides  the minimum of the action and never crosses a switching line. An instanton is a particular member of the minimum-action family of paths  that connects the two maxima of the potential $-V$.  It corresponds to  the  heteroclinic orbit    (${\bf x}^{*}(\tau)$ ,${\bf p}^{*}(\tau)$)  contained in the unstable Lagrangian manifold shared by the two maxima Ref.~\onlinecite{PhysRevA.65.032122}.
This explains why ground state tunneling splitting  for the particle in a multidimensional  potential is always described  by the instanton with a real-valued action and therefore can be simulated efficiently by PIMC.
 
This confirms that PIMC simulation of QMT in the ground state can be done without any loss of efficiency compared to what real system would do. The fact that PIMC simulations have the same scaling with the problem size as physical quantum annealing was recently experimentally confirmed, again on a spin system on chimera graph\cite{denchev2015computational}.
In this context,  it is  unlikely that  QA to find a ground state of optimization problem can achieve an exponential speedup over classical computation, only by using QMT as a computational resource.

We remark here that these conclusions hold only when  so-called \emph{stoquastic} Hamiltonians are used, i.e.  Hamiltonians which allow PIMC simulations. This is the case of the Hamiltonians used in this paper.
In most reaction simulations protons are assumed to be distinguishable particles, and, in QA the standard transverse field Ising Hamiltonian is also \emph{stoquastic}.\cite{johnson2011quantum,dwave2x}
This provides additional evidences that QA machines should implement \emph{non-stoquastic}
Hamiltonians which display sign-problem,\cite{PhysRevE.85.051112,mazzola2017quantum,hormozi2016non} to avoid efficient simulations by QMC methods.

{\bf Acknowledgements.}
GM acknowledges useful discussions with P. Faccioli, S. Sorella and G.E. Santoro.
MT acknowledges hospitality of the Aspen Center for Physics, supported by NSF grant PHY-1066293.
Simulations were performed on resources provided  by the Swiss National Supercomputing Centre (CSCS).
This work was supported by the European Research Council through ERC
Advanced Grant SIMCOFE by the Swiss National Science Foundation through
NCCR QSIT and NCCR Marvel. This paper is based upon work
supported in part by ODNI, IARPA via MIT Lincoln Laboratory Air Force
Contract No. FA8721-05-C-0002. The views and conclusions contained
herein are those of the authors and should not be interpreted as necessarily
representing the official policies or endorsements, either expressed
or implied, of ODNI, IARPA, or the U.S. Government. The U.S. Government
is authorized to reproduce and distribute reprints for Governmental
purpose not-withstanding any copyright annotation thereon.


%



\end{document}